\documentclass[11pt]{article}

\usepackage{authblk}
\usepackage{float}
\usepackage{graphicx}
\usepackage{caption}
\usepackage{subcaption}
\usepackage{url}
\usepackage{multirow}
\usepackage{titlesec}
\usepackage{setspace}
\usepackage{threeparttable}
\usepackage{amsmath}
\usepackage{amsfonts}
\usepackage{url}
\usepackage[numbers,square,sort&compress]{natbib}
\usepackage{booktabs}

\title{Algorithmic Intervention Decreases Polarization in Content Moderation}

\author[1]{Anna Bertani\thanks{Corresponding author}}
\author[1,2]{Saeedeh Mohammadi}
\author[1,2]{Taha Yasseri}

\affil[1]{Centre for Sociology of Humans and Machines, Trinity College Dublin and Technological University Dublin  \\ Dublin D02 PN40, Ireland}
\affil[2]{School of Mathematics and Statistics, University College Dublin \\ Dublin D04 V1W8, Ireland}
\date{}
\begin{document}
\maketitle

\doublespacing
\abstract{Crowd-based moderation relies on people's collective judgments to combat misinformation, but raters bring their own ideological biases to the task, shaping how they moderate each other's contributions. We ask whether an algorithmic redesign can counteract this bias, using X's Community Notes, where contributors write context for potentially misleading posts and rate each other's notes. Community Notes replaced simple majority voting with an algorithm that instead rewards agreement across people with different ideological viewpoints. Comparing 8{,}805 notes and 4{,}367 contributors before and after this change, we find that ideological polarization decreased in the algorithm’s estimates of contributor and note positions. Although negative interactions increased, contributors became less structurally aligned, indicating that negative interactions were less organized along partisan lines. These findings show that algorithmic incentives can reshape both the social structure and behavioral dynamics of crowds, extending beyond their intended effects on content moderation.}


\maketitle

\section{Introduction}\label{sec1}

Social media algorithms are often blamed for amplifying the societal harms of social media. By prioritizing rapid information diffusion \cite{vosoughi2018spread} and personalizing content based on user behavior \cite{lazer2018science}, these algorithms have been linked to the spread of misinformation, political polarization, and the erosion of public discourse \cite{klein2025mobilising}, with consequences for public health \cite{loomba2021measuring, enders2022relationship} and democratic processes \cite{allcott2017social, greene2021misremembering}. But what happens when an algorithm is designed to reduce polarization rather than amplify it? This question is particularly important for content moderation systems, where platforms must respond to harmful content at the speed and scale at which it spreads, without sacrificing accuracy, user trust, or freedom of expression.

Social media platforms have relied on professional fact-checkers to fight misinformation. This approach has proven effective: fact-checking can reduce the spread of false information and improve the accuracy of users' beliefs \cite{fazio2024combating, hoes2024prominent}. However, its limitations are equally well documented. Professional fact-checking is labor-intensive, operates slowly relative to the speed at which content can go viral, and is difficult to scale to the volume of claims circulating daily on large platforms \cite{hassan2015quest, nakov2021automated}. Fact-checking has also faced criticism regarding potential political and cultural bias. A 2019 survey found that approximately half of Americans believed that fact-checkers were biased, with this perception being substantially more prevalent among Republicans than Democrats \citep{walker2019factcheckers}. Such perceptions can undermine the perceived legitimacy of moderation systems.

In response to these limitations, several major platforms have begun shifting from expert-driven moderation towards community-based systems, in which users themselves can act as fact-checkers by collectively evaluating and annotating potentially misleading content. A prominent example of such initiatives is Community Notes (Birdwatch), which was first introduced on X (Twitter) \cite{twitter2021birdwatch} and is currently being implemented by Meta \cite{meta2025}. In Community Notes, contributors write notes to add missing context or correct potentially misleading information. Other contributors rate the helpfulness of these notes, and an algorithm uses these ratings to determine which notes are displayed to users \cite{communityNotesGettingStarted}.

Early evidence suggests that Community Notes can achieve levels of accuracy comparable to those of professional fact-checkers \cite{allen2021scaling, la2024crowdsourced}, while offering a more scalable and cost-effective alternative to expert-based models \cite{saeed2022crowdsourced, martel2024crowds}. Community-based notes may also be more persuasive because users tend to prefer contextual explanations over authoritative declarations of truth \cite{drolsbach2024community}. Moreover, Community Notes are a less intrusive intervention, as users retain the autonomy to engage with or disregard the information in the notes \cite{augenstein2025community}.

Displayed notes also reduce engagement with misleading posts, which receive fewer reposts and are more likely to be deleted \cite{chuai2026community}. However, due to the sparsity of contributor ratings, only approximately 13\% of posts receiving at least one note have a displayed note. Moreover, displayed notes appear, on average, 26 hours after the original post \cite{mohammadi2025birdwatch}, while the half-life of reposts on X is approximately 6 hours \cite{chuai2026community}. Notes often become visible long after much of the content's circulation has occurred. Both limited coverage and delayed visibility can substantially reduce the effectiveness of moderation interventions \cite{truong2025delayed}.

Originally, Notes surfaced based on a majority rule, in which Notes were displayed when they received a sufficiently high proportion of positive ratings. Contributors tend to rate notes written by ideologically aligned contributors as Helpful and those written by opposing contributors as Not Helpful \citep{allen2022birds}. This pattern was also reflected in contributor interaction networks, which exhibited polarization similar to that observed in social media friendship networks \citep{yasseri2023can}, making the system vulnerable to partisan coordination, bias, and subjectivity.

To address this problem, X replaced the majority rule with a bridging algorithm \cite{wojcik2022birdwatch}. The algorithm uses historical rating patterns to estimate latent positions for contributors and notes, which are then used to weight ratings when calculating a note’s helpfulness. Consequently,  positive ratings from contributors who have disagreed with each other in the past, carry more weight than the ones who usually agree. The redesign therefore shifted the aggregation mechanism from relying primarily on the proportion of positive ratings toward accounting for the diversity of rating patterns.

The emphasis on the cross-ideological consensus may itself contribute to determining which notes are able to become visible; posts that are more polarizing appear less likely to receive notes that achieve such consensus \cite{bouchaud2026community}.

These limitations raise a broader question: how does the bridging algorithm shape the behavior of the people who use them? Did the introduction of bridging-based aggregation change patterns of contributor behavior, and, if so, how did these changes manifest in the system’s latent ideological structure and patterns of agreement? More importantly, did the redesign improve the quality of notes?

Here, we compare Community Notes contributors and notes under the majority rule and the bridging algorithm, using matched samples with comparable numbers of contributors, notes, and ratings over similar observation windows. The new algorithm is designed to select better notes, but the more important question is whether it also changes contributor behavior. Because many of the same contributors wrote and rated notes both before and after its introduction, this comparison lets us ask whether their behavior itself changed, rather than only which notes the algorithm now favors. This offers a quasi-experimental opportunity \cite{hale2018digital, ternovski2020social, yasseri2022fooled}, though, as with any such design, it cannot fully rule out concurrent changes in the platform, its user population, or the broader information environment. For each period, we estimate the latent polarity parameters generated by the matrix factorization model and construct signed contributor interaction networks from rating behavior, capturing both the algorithmic and social dimensions of note evaluation.

\section{Results}

\begin{figure}[H]
\centering
\includegraphics[width=0.85\textwidth]{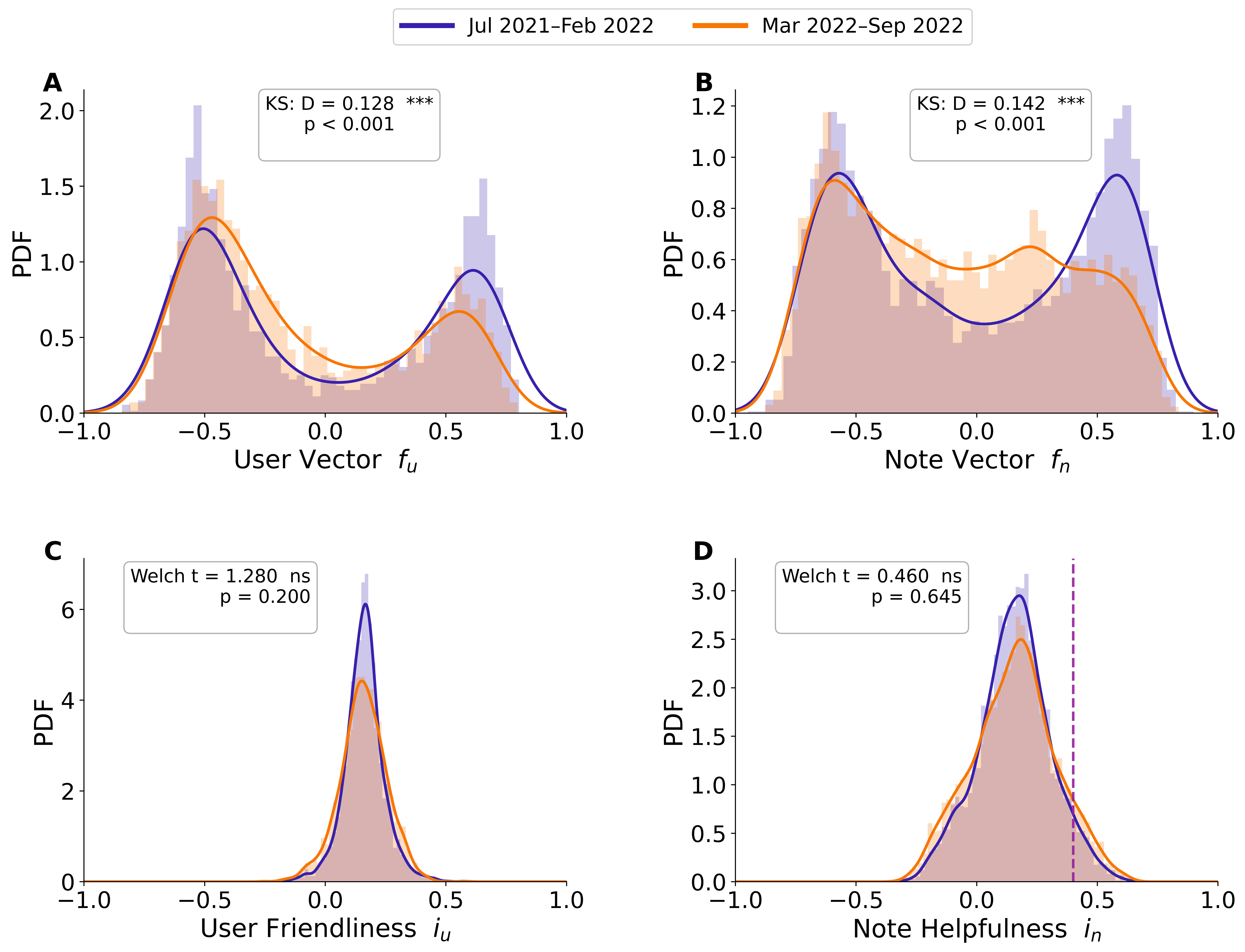}
\caption{\textbf{Estimated distributions of Community Notes Matrix Factorization parameters before and after March 2022.} Panel A shows Contributor Vector ($f_u$), the inferred ideological position of raters (KS: $D = 0.128$, $p < 0.001$). Panel B shows Note Vector ($f_n$), the ideological position of notes (KS: $D = 0.142$, $p < 0.001$). Panel C shows User Friendliness ($i_u$), the per-rater tendency to vote Helpful (Welch $t = 1.280$, $p = 0.200$). Panel D shows Note Helpfulness ($i_n$), the estimated quality of notes; the dashed line marks the publication threshold ($i_n \geq 0.40$) (Welch $t = 0.460$, $p = 0.645$).}\label{fig:distributions}
\end{figure}

Figure~\ref{fig:distributions} compares the two periods before  and after the introduction of the bridging algorithm across four measures: the latent polarity of contributors and notes, contributor leniency in assigning Helpful ratings, and note helpfulness after accounting for ideological polarity (Materials and Methods). Contributors and Notes became less ideologically divided under the bridging algorithm, while contributor leniency and note quality stayed the same. Both ideological positions showed a bimodal distribution that became less pronounced after the transition (Kolmogorov–Smirnov test: contributors, $D = 0.128$, $p < 0.001$; notes, $D = 0.142$, $p < 0.001$). By contrast, neither contributor leniency nor note helpfulness changed significantly (Welch’s $t = 1.280$, $p = 0.200$, and $t = 0.460$, $p = 0.645$, respectively). One might suspect that this difference is given by the different composition of the contributor population. When we restricted the analysis to contributors active in both periods (Fixed Pool Users Sample), the polarity distributions remained significantly different (see Supplementary Information Fig. \ref{fig:sensitivity_distribution}).

\begin{figure}[H]
\centering
\includegraphics[width=0.85\textwidth]{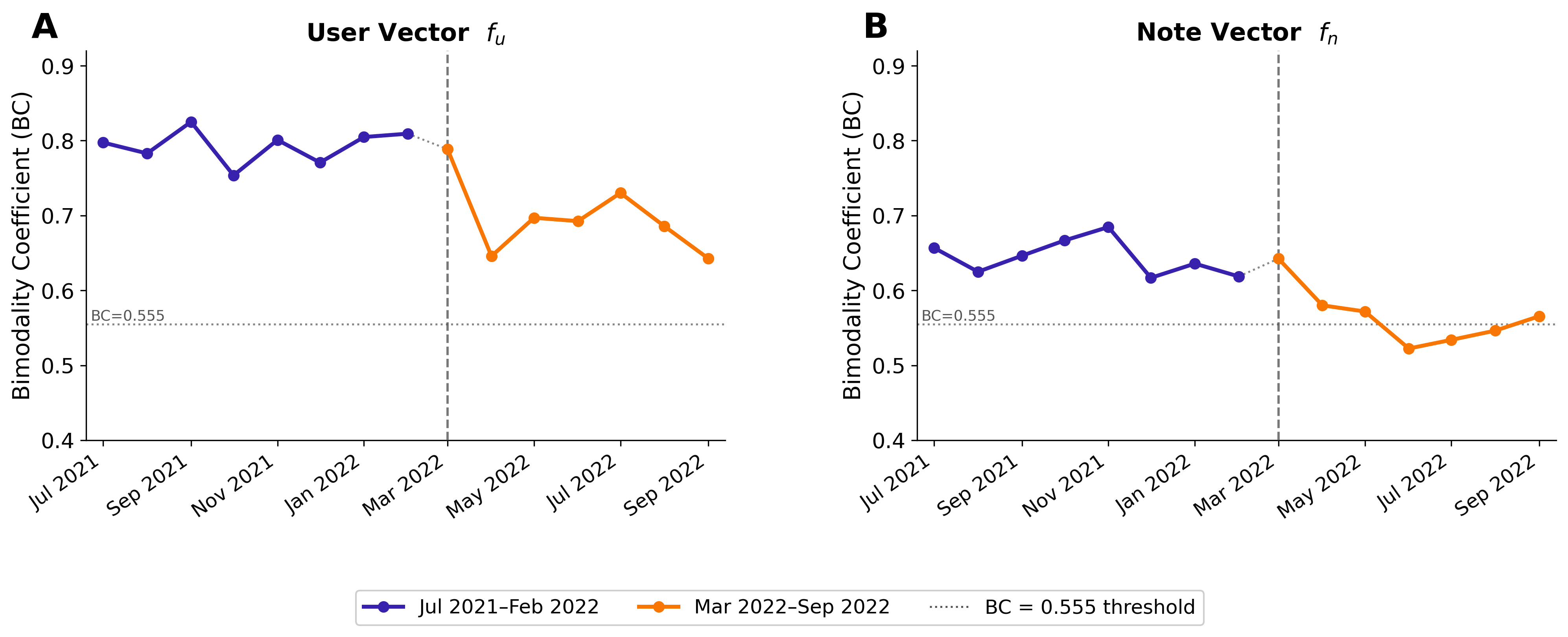}
\caption{\textbf{Monthly evolution of Community Notes Matrix Factorization parameters before and after March 2022.} Panel A shows the BC of the Contributor Vector ($f_u$), which remains consistently above the bimodality threshold ($\text{BC} = 0.555$, dashed grey line) throughout both periods but declines sharply after March 2022, indicating weaker rater polarization. Panel B shows the BC of Note Vector ($f_n$), which follows a similar declining trajectory and approaches the unimodal threshold following the introduction of the bridging algorithm.}\label{fig:trajectories}
\end{figure}

To track how polarization declined over time, we fitted the model month by month and measured the bimodality of $f_u$ and $f_n$ using the Bimodality Coefficient (BC), which exceeds 0.555 for bimodal distributions (Materials and Methods). Figure~\ref{fig:trajectories} shows that BC declined for both $f_u$ (Panel A) and $f_n$ (Panel B) starting in March 2022, approaching the bimodality threshold by the end of the observation window, with the decline more pronounced for contributors than for notes. This suggests that the ideologically bimodal structure the algorithm infers from contributor and note behavior weakened gradually, consistent with a gradually more heterogeneous contributor population rather than an abrupt shift.

\begin{figure}[H]
\centering
\includegraphics[width=\textwidth]{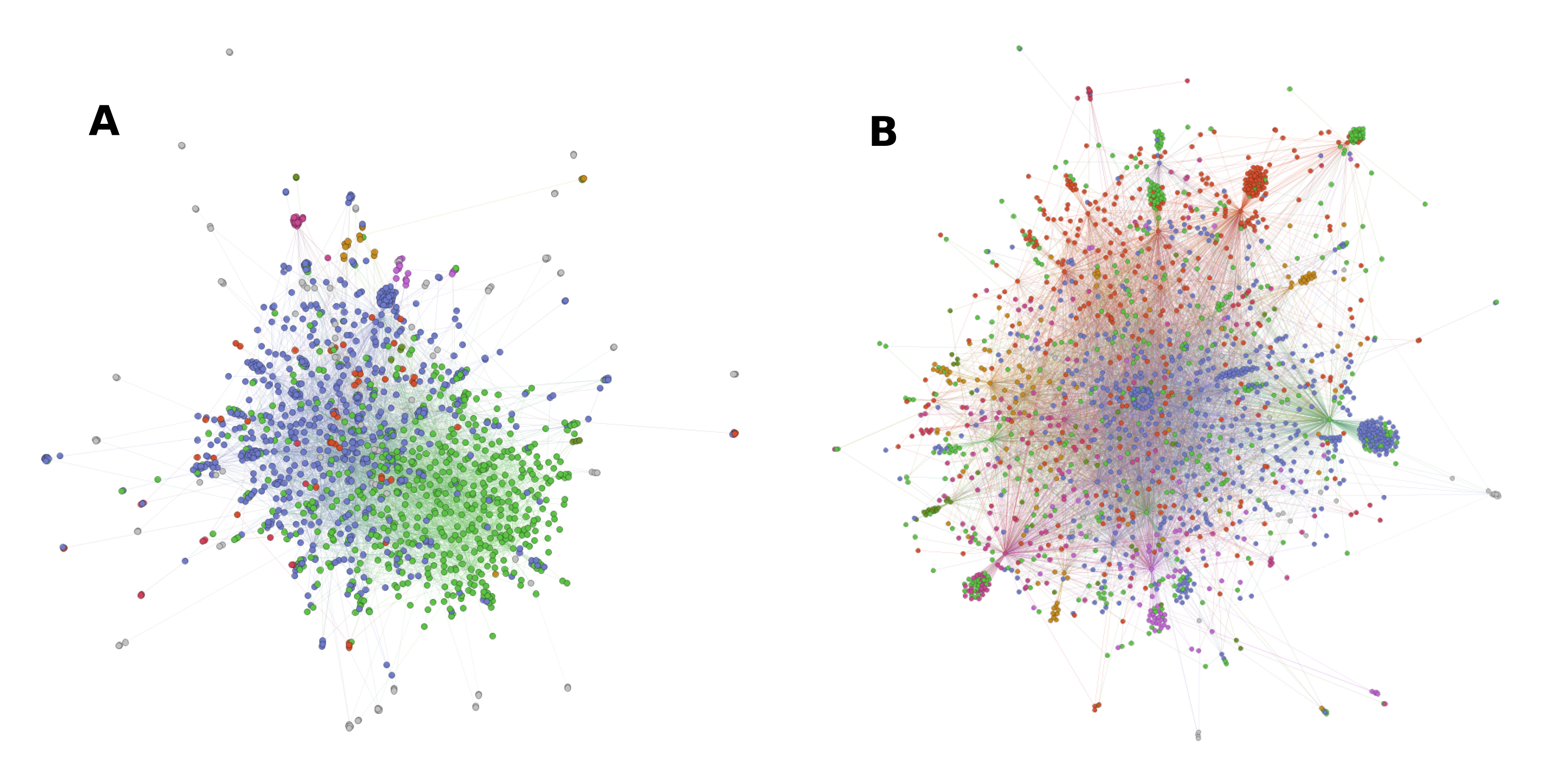}
\caption{\textbf{Signed interaction networks before and after the implementation of the bridging algorithm.} 
Panel A shows the signed directed network for the first period, and Panel B shows the corresponding network for the second one. Nodes represent Community Notes contributors, and node colors indicate detected communities. Networks are visualized using the ForceAtlas2 layout with LinLog mode \cite{jacomy2014forceatlas2}; edge signs are not used in the layout.}
\label{fig:network_vis}
\end{figure}

Having shown that contributors' latent ideology became less bimodal over time, we next asked whether this was also visible in how contributors actually interacted with one another, independent of the model. To study these interactions, we built a signed, directed network for each period from contributor ratings (see Materials and Methods for more details). Figure~\ref{fig:network_vis} shows the networks. The first network separates into two large, distinct clusters (modularity = 0.85), while the second one is characterized by higher fragmentation, with more communities (modularity = 0.56).

To further examine changes in network structure, we calculated the alignment, how cleanly contributors split into two opposing groups, and antagonism, the share of interactions that are negative, before and after the introduction of the bridging algorithm (see Materials and Methods). The alignment score decreases (two-sided $z$-test $z =  159.86$, $P<0.001$) from 0.740 (95\% CI [0.739, 0.741]) before the transition
to 0.649 (95\% CI [0.648, 0.650]) after it,  
demonstrating that contributor interactions become less cleanly organized into structurally balanced opposing groups following the introduction of the bridging algorithm. At the same time, the antagonism score increases (two-sided $z$-test $z =  -8.28$, $P<0.001$) from 0.320 
 to 0.390 
 showing a higher prevalence of negative interactions in the second period.

This pattern is robust to alternative variance thresholds used in the network-construction procedure. Sensitivity analyses across these thresholds are reported in the Supplementary Information (Figure~\ref{fig:sensitivity_antagonism_alignment}).The pattern also holds when we restrict the analysis of the After period to contributors who were also present in the Before period (see Supplementary Information, Section B, for details).

To examine whether these differences are also visible over time, we additionally calculated alignment and antagonism for monthly contributor interaction networks (see Supplementary Information \ref{fig:alignment_antagonism_monthly})

\begin{figure}[H]
\centering
\includegraphics[width=.75\textwidth]{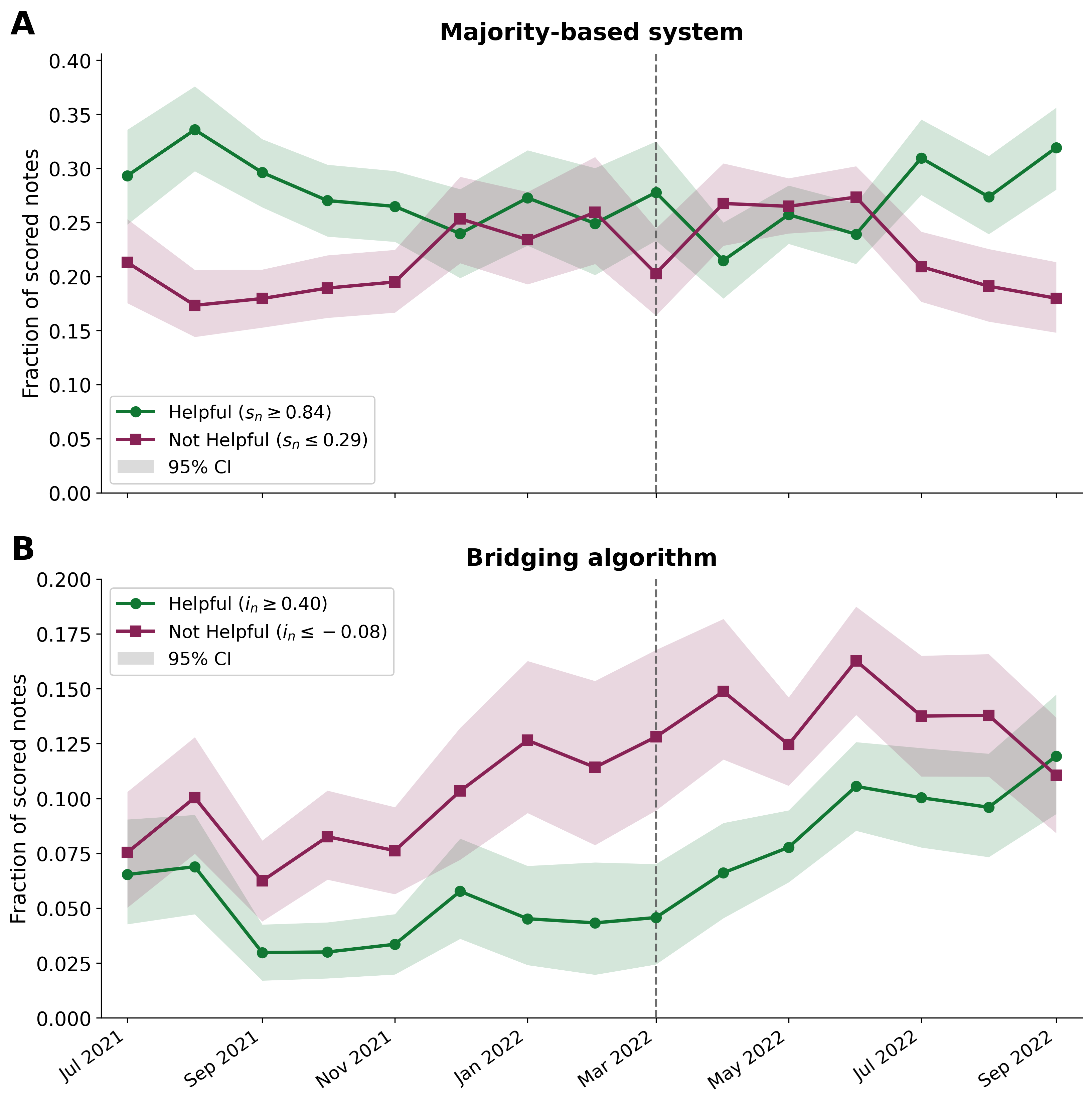}
\caption{\textbf{Comparing majority and bridging algorithm classification of note helpfulness over time.} Each panel shows the monthly fraction of scored Notes classified as Helpful or Not Helpful, split into the two periods before and after the February 2022 change. Shaded bands show 95\% confidence intervals ($\pm 1.96 \times \mathrm{SE}$), computed from the standard error of a proportion, $\mathrm{SE} = \sqrt{p(1-p)/n}$, where $p$ is the monthly rate and $n$ is the number of scored notes that month. (A) Majority rule: a note is Helpful if $\geq 84\%$ of raters mark it helpful, Not Helpful if $\leq 29\%$. (B) Bridging algorithm: a note is Helpful if its fitted intercept $i_n \geq 0.40$, Not Helpful if $i_n \leq -0.08$.}\label{fig:pubrates}
\end{figure}

We next examine whether these changes were accompanied by shifts in note classifications and, separately, whether notes classified as Helpful were judged to be more helpful by independent evaluators.
Figure~\ref{fig:pubrates} shows the monthly proportions of notes classified as Helpful and Not Helpful under the majority rule (Panel A) and the bridging algorithm (Panel B). Under the first algorithm, neither classification shows a clear trend after the transition. Under the second one, both Helpful and Not Helpful rates increase, although the overall share of classified notes remains lower than under the majority rule. Thus, despite classifying fewer notes overall, the second algorithm becomes more decisive over time, assigning a status to a growing proportion of notes.

Beyond the overall status of the notes classified, we asked whether the new algorithm selected better notes. To test this, we randomly sampled 160 notes, 80 in each period, and asked 10 independent evaluators to judge whether each note was helpful after reading the post and the note (Materials and Methods). Notes posted after the transition scored higher on average than those posted before (mean: 28.8 vs. 15.5; median: 40 vs. 20), though this difference was not statistically significant (Mann-Whitney $U = 2786.5$, $p = 0.155$). Still, a higher share of notes scored nonnegative, and fewer scored negative, after the transition than before (see Supplementary Information).

\section{Discussion}\label{sec12}

Community Notes’ shift to a bridging-based algorithm that rewards agreement among contributors who usually disagree allows us to examine two questions: does it change contributor behavior, and does it improve the quality of the resulting classifications? Our findings suggest that the transition was associated with a less ideologically structured pattern of participation and higher note helpfulness. At the same time, negative interactions among contributors increased.

We find evidence that contributor behavior became less structured along ideological lines following the introduction of the bridging algorithm. The polarity factors for both contributors and notes became less bimodal, with this change accumulating over time rather than emerging as an abrupt shift. This gradual pattern is consistent with behavioral adaptation to the new incentive structure. Previous research showed that the original system exhibited substantial partisan clustering, with contributors tending to support notes from ideologically similar contributors and reject those from opponents \cite{allen2022birds}. 
Under the majority-based system, notes could receive a label when supported by a sufficiently large share of contributors, even if that support was concentrated within a particular ideological group. The bridging algorithm, in contrast, gives greater weight to agreement among contributors who tend to disagree with one another. This change may encourage contributors to evaluate notes beyond their usual ideological boundaries, making cross-group agreement more important for determining which notes receive a label. Our findings are consistent with \cite{wojcik2022birdwatch}, who argued that bridging-based aggregation mechanisms could reduce the influence of partisan majorities and elevate content that can receive cross-group support.

Based on the outcomes of the bridging algorithm, both the Helpful and Not Helpful classifications increased after the transition. Also, the Helpful rate increased from a much lower base, narrowing the gap between them over time. This pattern is consistent with an expanding capacity to achieve cross-ideological consensus, not only in identifying unhelpful content but also in recognizing helpful contributions. This difference in how many notes each algorithm resolves highlights a broader trade-off in bridging-based moderation. By weighting agreement across ideological groups more heavily than a simple majority, the algorithm uses a more conservative threshold for assigning labels, leaving more notes unresolved but requiring broader agreement for those it classifies. The system therefore trades coverage for stronger agreement across ideological groups. Whether this is desirable depends on what the platform values the most: labeling more content or ensuring that the labels it assigns are supported by contributors from different parts of the ideological spectrum. Beyond the effect on classification rates, the transition may also have improved note quality. Based on external evaluations, we found an increase in notes considered more helpful following the implementation of the second algorithm. 

The rise in Not Helpful notes aligns with the increase in antagonism we observed in the second period: contributors evaluated one another more negatively after the transition. Yet this did not translate into stronger ideological clustering — alignment, the extent to which contributors organize into opposing groups \cite{fraxanet2024unpacking}, decreased over the same period. Contributors became less organized into coherent ideological clusters even as their interactions grew more negative overall. One possible explanation lies in how the platform incentivizes rating in the first place. New contributors must accumulate sufficient rating impact, which reflects how closely their ratings align with the consensus of the crowds, before they are allowed to write notes themselves \cite{xcorp2026writingimpact}. Rating a note as Not Helpful is often a lower-effort way to earn a rating impact: a Not Helpful rating can be justified by pointing to an obvious flaw, whereas a Helpful rating typically requires verifying that the note’s sources actually support its claim. This could explain both the rise in Not Helpful classifications and the increase in antagonism we observe, independent of any change in the underlying quality of notes.

Our results are subject to several limitations. First, the study relies on observational data surrounding a platform-level intervention and therefore cannot establish causality.  As with any natural experiment that exploits a real-world policy change rather than a randomized design, we lack full experimental control over which contributors were exposed to the new algorithm and when. Other simultaneous changes, including growth in the contributor population and broader platform developments, may also have influenced contributor behavior. However, we compared our results with the Fixed Pool Users Sample of contributors both before and after, and the results were consistent (see Supplementary Information). In addition, the change was explicitly communicated to contributors through a banner in the rating interface explaining the shift away from majority rule. This communication provides a plausible mechanism linking the change in the platform’s incentive structure to the behavioral shifts we observe. Second, the analysis focuses on English-language Community Notes. During the observation window, Community Notes operated primarily in English-language markets, so this scope captures most of the notes written in this period rather than a restricted subset; nonetheless, the long-term effects of the bridging algorithm may differ as the system has since expanded to additional countries and languages. 
Third, the latent ideological dimensions estimated by the Matrix Factorization model should not be interpreted as direct measures of political ideology. Rather, they represent behavioral dimensions inferred from rating patterns and may capture additional forms of disagreement beyond partisan identity. However, we sampled the notes in this evaluation without considering topic, and we do not examine topic differences in this study. Future research could investigate whether the accuracy of Helpful and Not Helpful classifications, as well as the level of independent agreement with these classifications, varies across topics. 

Despite these limitations, the study provides one of the first empirical evaluations of how a large-scale algorithmic redesign affects both moderation outcomes and contributor interactions within a community-based content moderation system. These results also contribute to broader discussions of algorithmic governance. Platform interventions are often evaluated in terms of their ability to reduce harmful behavior or improve informational quality. Our findings suggest one possible explanation is that the intervention reshaped the form of social conflict rather than simply eliminating it.

\section{Material and Methods}

\subsection{Data}

We use a corpus of English-language Community Notes (formerly Birdwatch) 
collected between July 2021 and September 2022 \cite{mohammadi2025birdwatch}, which contains each note, the post it refers to, and all ratings it received from other contributors. Each rating is one of three options: \texttt{HELPFUL}, \texttt{SOMEWHAT\_HELPFUL}, or \texttt{NOT\_HELPFUL}.

 On 28 February 2022, the platform replaced a simple majority rule with the bridging-based Matrix Factorization algorithm \cite{wojcik2022birdwatch}. The platform documented this change through a public announcement and, more directly, a banner shown to contributors within the rating interface explaining that note status would no longer be determined by majority rule (see Supplementary Information).
 We compare two periods around this date:

\begin{itemize}
    \item 1 July 2021 -- 28 February 2022: the period 
    before deploying the bridging algorithm.
    \item 1 March 2022 -- 30 September 2022: the period 
    immediately following the deployment of the bridging algorithm. 
\end{itemize}

\subsection{Filtering Pipeline}

Before estimating the Matrix Factorization parameters, the Community Notes algorithm applies a sequence of filters that reduce the number of notes and raters entering the model relative to the raw input \cite{communitynotes_contributor_scores}. First, a note must have received at least five ratings before a helpfulness label can be assigned. Second, a rater must have submitted at least ten ratings before their contributions are counted toward note scoring. Third, a ``helpfulness filter'' removes raters whose rating behavior diverges systematically from the group consensus. Table~\ref{tab:filtering} reports the number of notes and raters in the two windows before and after the algorithm change.

\begin{table}[h]
\centering
\caption{Number of notes and raters in the two periods, unfiltered (raw input) and filtered (scored output) \cite{communitynotes_contributor_scores}.}
\label{tab:filtering}
\begin{tabular}{lrrrr}
\toprule
 & \multicolumn{2}{c}{\textbf{Jul 2021--Feb 2022}} &
   \multicolumn{2}{c}{\textbf{Mar 2022--Sep 2022}} \\
\cmidrule(lr){2-3}\cmidrule(lr){4-5}
 & Notes & Raters & Notes & Raters \\
\midrule
Raw input     &  9{,}813 &  6{,}155 &  7{,}895 & 12{,}813 \\
Scored output &  4{,}139 &  2{,}196 &  4{,}666 &  2{,}171 \\
\bottomrule
\end{tabular}
\end{table}

\subsection{The Community Notes Bridging Algorithm}

For each rater $u$ and note $n$, the model predicts a rating $\hat{r}_{un}$, the predicted helpfulness of note $n$ to rater $u$, as:

\begin{equation}
    \hat{r}_{un} = \mu + i_u + i_n + f_u \cdot f_n
\end{equation}

\noindent where $\mu$ is a global intercept representing the General Mood, $i_u$ is the Contributor Friendliness (per-rater tendency to vote Helpful), and $i_n$ is the Note Helpfulness (the latent quality of the Note). $f_u$, the Contributor Vector, and $f_n$, the Note Vector, are latent vectors representing rater $u$'s and note $n$'s position in an inferred space; their dot product, $f_u \cdot f_n$, is an interaction term that grows larger when rater and note occupy similar positions, capturing the ideological alignment between them. The model is estimated via Stochastic Gradient Descent with $\ell_2$ regularization, using regularization coefficients $\lambda_i = 0.15$ for the intercept terms ($i_u$, $i_n$) and $\lambda_f = 0.03$ for the polarity factors ($f_u$, $f_n$). A note receives the label \textit{Currently Rated Helpful} if $i_n \geq 0.40$, \textit{Currently Rated Not Helpful} if $i_n \leq -0.08$, and \textit{Needs More Ratings} otherwise.

We run the algorithm using the open-source implementation from \cite{wojcik2022birdwatch}, replicating the scoring pipeline using only the Core MF scorer.

\subsection{Statistical Tests}

We quantify the degree of bimodality in a distribution using the Bimodality Coefficient:

\begin{equation}
    \text{BC} = \frac{\text{skewness}^2 + 1}{\text{kurtosis}_{\text{Pearson}}}
\end{equation}

\noindent where $\text{BC} > 0.555$ indicates bimodality. We use this coefficient to decide, for each parameter, which test to apply: for distributions found to be bimodal, we compare the two periods using the two-sample Kolmogorov--Smirnov (KS) test, which makes no distributional assumptions; for distributions that are approximately unimodal and symmetric, we use Welch's $t$-test, which does not assume equal variances.

\subsection{Helpful and Not Helpful Classification}

For each algorithm, we compute the monthly share of notes classified as Helpful and as Not Helpful. Under the majority rule, a note is classified Helpful if at least 84\% of its ratings are positive, and Not Helpful if at most 29\% are positive \cite{crisman2021birdwatch}. Under the Matrix Factorization rule, a note is classified Helpful if $i_n \geq 0.40$, and Not Helpful if $i_n \leq -0.08$. All other notes remain unclassified (\textit{Needs More Ratings}).

\subsection{Evaluation of the helpfulness of notes}

We conducted a crowd-based evaluation to measure agreement among independent raters on the perceived helpfulness of the notes attached to the posts.

Participants were recruited via Prolific and redirected to an experimental interface implemented in oTree. After providing informed consent, each participant was randomly assigned to evaluate a single note.

The stimulus set consists of 160 community notes sampled from a larger corpus: 80 drawn at random from each period considered, without reference to their Helpful or Not Helpful classification. For each note, participants were shown a screenshot of the corresponding post together with the associated note.

Each note received 10 independent evaluations (plus one attention check), for a total of 1{,}600 observations. Participants were asked: \textit{``Do you think that the above text or note is helpful or not helpful with respect to the post shown above?''} Responses were recorded as a binary choice (Helpful / Not helpful) (see Supplementary Information for details).

\subsection{Contributor Interaction Networks}

We construct separate signed, directed networks for both periods. Nodes represent contributors. A directed edge from contributor $i$ to contributor $j$ indicates that contributor $i$ rated at least one Note written by contributor $j$ during the corresponding period. Thus, the source node is the rater and the target node is the note author.

Following \cite{fraxanet2024unpacking}, we model the relationship from rater $i$ to note author $j$ as a Bernoulli random variable with probability $p_{ij}$. We estimate $p_{ij}$ using a Bayesian beta-binomial model with a uniform prior, $\alpha_0 = 1$ and $\beta_0 = 1$.

For each ordered pair $(i,j)$, we aggregate all ratings that contributor $i$ assigned to notes written by contributor $j$. Helpful ratings are treated as positive evidence, Not Helpful ratings as negative evidence, and Somewhat Helpful ratings as partial positive evidence. The posterior parameters are therefore defined as:

\begin{equation}
    \alpha_{ij} = \alpha_0 + H_{ij} + 0.5 \times S_{ij}
\end{equation}

\begin{equation}
    \beta_{ij} = \beta_0 + N_{ij}
\end{equation}

\noindent where $H_{ij}$ is the number of Helpful ratings, $S_{ij}$ is the number of Somewhat Helpful ratings, and $N_{ij}$ is the number of Not Helpful ratings that contributor $i$ assigned to notes written by contributor $j$.

The posterior expectation and variance of $p_{ij}$ are:

\begin{equation}
    \mathbb{E}[p_{ij}] = \frac{\alpha_{ij}}{\alpha_{ij} + \beta_{ij}}
\end{equation}

\begin{equation}
    Var[p_{ij}] = \frac{\alpha_{ij}\beta_{ij}}
    {(\alpha_{ij} + \beta_{ij})^2(\alpha_{ij} + \beta_{ij} + 1)}
\end{equation}

We retain an edge only when the estimated relationship is both sufficiently polarized and sufficiently certain. Specifically, we retain a positive edge when $\mathbb{E}[p_{ij}] \geq 0.6$ and $Var[p_{ij}] \leq 0.05$, and a negative edge when $\mathbb{E}[p_{ij}] < 0.4$ and $Var[p_{ij}] \leq 0.05$. Positive edges, therefore, indicate significant positive interaction, while negative edges indicate significant negative interaction. Pairs with intermediate expected values or high posterior uncertainty are excluded from the signed network.

To assess the robustness of our network construction to this uncertainty threshold, we conducted a sensitivity analysis using variance thresholds ranging from 0.01 to 0.1 in increments of 0.005. The results are reported in the Supplementary Information. 

Using this procedure, the first network contains 1,464 nodes and 4,881 signed edges, constructed from 91,006 raw rating interactions. The second network contains 1,846 nodes and 6,933 signed edges, constructed from 101,394 raw rating interactions. Figure~\ref{fig:network_vis} visualizes the resulting signed networks.

\subsection{Network Antagonism and Alignment}
Following \cite{fraxanet2024unpacking}, we distinguish between two dimensions of polarization in signed networks: antagonism and alignment. Antagonism captures the prevalence of negative interactions in the network, whereas alignment captures how consistently positive and negative interactions are organized across groups.

\paragraph{Antagonism}
Antagonism is the percentage of negative edges among all edges in the signed network.

\paragraph{Alignment}
In a signed network, an edge is considered frustrated if it is negative within a community or positive between communities. The frustration score is the total number of frustrated edges in a two-community partition of the network.

For each signed network, we first approximate the minimum frustration score using simulated annealing. Nodes are randomly assigned to one of two communities, and the algorithm iteratively proposes moving a single node to the other community. We always accept a proposed move if it reduces the number of frustrated edges; otherwise, we accept it with a probability that decreases over the annealing schedule. We use 200,000 annealing steps per run and repeat the procedure across 200 random restarts. We retain the lowest frustration score obtained across all restarts as the observed frustration score, $L_{\text{obs}}$.

Because the observed level of frustration depends in part on the network's size and density, we compare the observed frustration score against a sign-randomized null model. For each network, we generate 100 null networks by randomly permuting the signs of the observed edges while preserving the underlying network topology and the overall number of positive and negative edges. We then estimate the minimum frustration score for each null network using the same simulated-annealing procedure. The mean frustration score across these null networks is denoted $L_{\text{null}}$.

We define the network's alignment as the proportional reduction in frustration relative to the sign-randomized null model:

\begin{equation}
    P = 1 - \frac{L_{\text{obs}}}{\overline{L}_{\text{null}}}
\end{equation}

\noindent where $L_{\text{obs}}$ is the observed minimum frustration score and $\overline{L}_{\text{null}}$ is the mean minimum frustration score across the sign-randomized null networks. Higher values indicate that the observed signed network is more structurally balanced, and therefore more polarized, than would be expected from a network with the same topology and sign distribution but randomly assigned edge signs.

To quantify uncertainty in the alignment scores, we bootstrap the null frustration distribution. We generate 2,000 bootstrap samples by repeatedly resampling the 100 null frustration scores with replacement. For each bootstrap sample, we calculate the mean null frustration score and use it to recalculate the alignment score. We report the resulting 95\% bootstrap confidence intervals.

\subsection{Community Detection}
For the network visualizations in Figure~\ref{fig:network_vis}, we used \textit{SignedLouvain} to detect communities. \textit{SignedLouvain} is a Louvain-based method for signed networks that optimizes signed modularity \cite{pougue2024signedlouvain}.
We represented the signed network as two layers: a positive-edge layer and a negative-edge layer. We used the default SignedLouvain layer weights, resolution parameters, and two-hop community search setting (layer\_weights = [1, -1], resolutions = [1, 1], k = 2). We set the maximum number of passes per level to 40.

\section*{Acknowledgments}
This publication has emanated from research supported in part by a grant from Taighde Éireann – Research Ireland under Grant numbers 18/CRT/6049 and IRCLA/2022/3217. TY acknowledges support from Workday, Inc.
During the preparation of this work, the authors used ChatGPT 5.6 to improve the writing style of this article. After using this tool, the authors reviewed and edited the content as needed and take full responsibility for the content of the publication.


\section*{Author Contribution}
AB and SM curated and analyzed the data. AB analyzed the algorithm and produced the corresponding visualizations. SM conducted the network analysis and produced the corresponding visualizations. AB and SM drafted the paper. AB, SM, and TY developed the paper's core ideas. TY supervised the project. All authors contributed to the writing of the manuscript and approved the final version for publication.

\section*{Competing interests}
The authors declare no competing interests.

\section*{Data and Material Availability}
Data are available at \url{https://doi.org/10.5281/zenodo.20591253}.

\bibliographystyle{unsrtnat}
\bibliography{sn-bibliography}
\renewcommand{\thetable}{S\arabic{table}}
\renewcommand{\thefigure}{S\arabic{figure}}

\setcounter{table}{0}  
\setcounter{figure}{0} 
\appendix
\newpage
\section*{SUPPLEMENTARY INFORMATION}

{\Large Algorithmic Intervention Decreases Polarization in Content Moderation}
\vspace{.4 cm}

{\large Anna Bertani, Saeedeh Mohammadi, Taha Yasseri}
\vspace{.4 cm}

\section{Introduction of the Bridging Algorithm}

\begin{figure}[H]
\centering
\includegraphics[width=\textwidth]{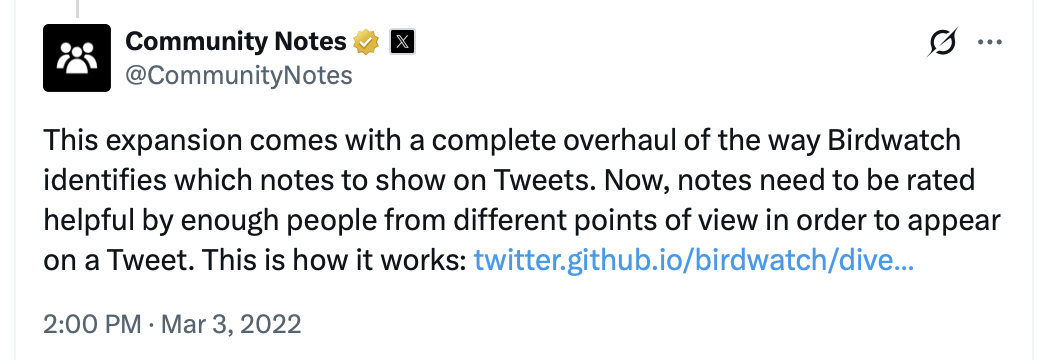}
\caption{\textbf{Public announcement of the Bridging Algorithm.} On March 3, 2022, Community Notes publicly announced the launch of the Bridging Algorithm on Twitter/X, describing the shift away from simple majority-rule scoring toward a system that gives greater weight to agreement between raters with different points of view. This announcement was visible to the platform's general public, independent of individual contributors' engagement with the Community Notes program.}
\label{fig:communitynote_launched}
\end{figure}

\begin{figure}[H]
\centering
\includegraphics[width=\textwidth]{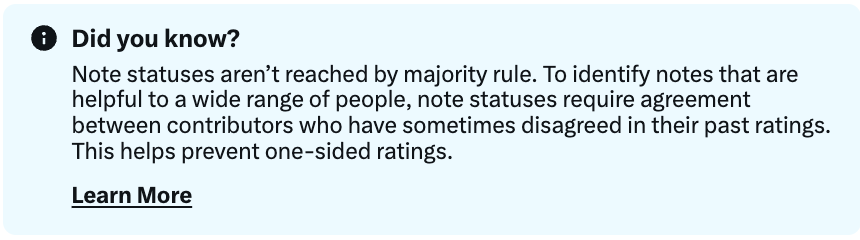}
\caption{\textbf{Banner announcing the new rating logic to contributors.} Banner shown to anyone with a Community Notes account within the rating interface at the time of the transition, informing them that note helpfulness would no longer be determined by majority rule. This banner reached contributors directly within their rating workflow, making it the more direct evidence that raters were exposed to a description of the new scoring rule.}
\label{fig:banner_message}
\end{figure}

With the introduction of the Bridging Algorithm in early March 2022, Community Notes replaced the previous majority rule for determining whether a note is Helpful. The platform documented this transition through two channels reaching different audiences: a public announcement on Twitter/X (Fig.~\ref{fig:communitynote_launched}), and a banner shown directly to raters within the rating interface (Fig.~\ref{fig:banner_message}). The algorithmic change was clearly documented both to the platform's general audience and to the specific population of contributors whose rating behavior we analyze in this paper. The in-product banner is the more consequential of the two for our identification strategy, since it is the channel most likely to have reached contributors directly within the workflow captured in our data.

\section{Robustness check on the same sample of contributors}

A potential concern with comparing the two periods is that the contributor population is not identical across the two windows: the number of raw raters rises from 6,155 to 12,813(see \ref{tab:filtering}). Any of our reported differences could therefore reflect compositional turnover in who participates, rather than existing contributors changing how they rate. To isolate the latter, we restrict the sample to the Fixed Pool Users Sample, contributors who appear as raters in both windows and repeat the two central analyses.

\begin{figure}[H]
\centering
\includegraphics[width=\textwidth]{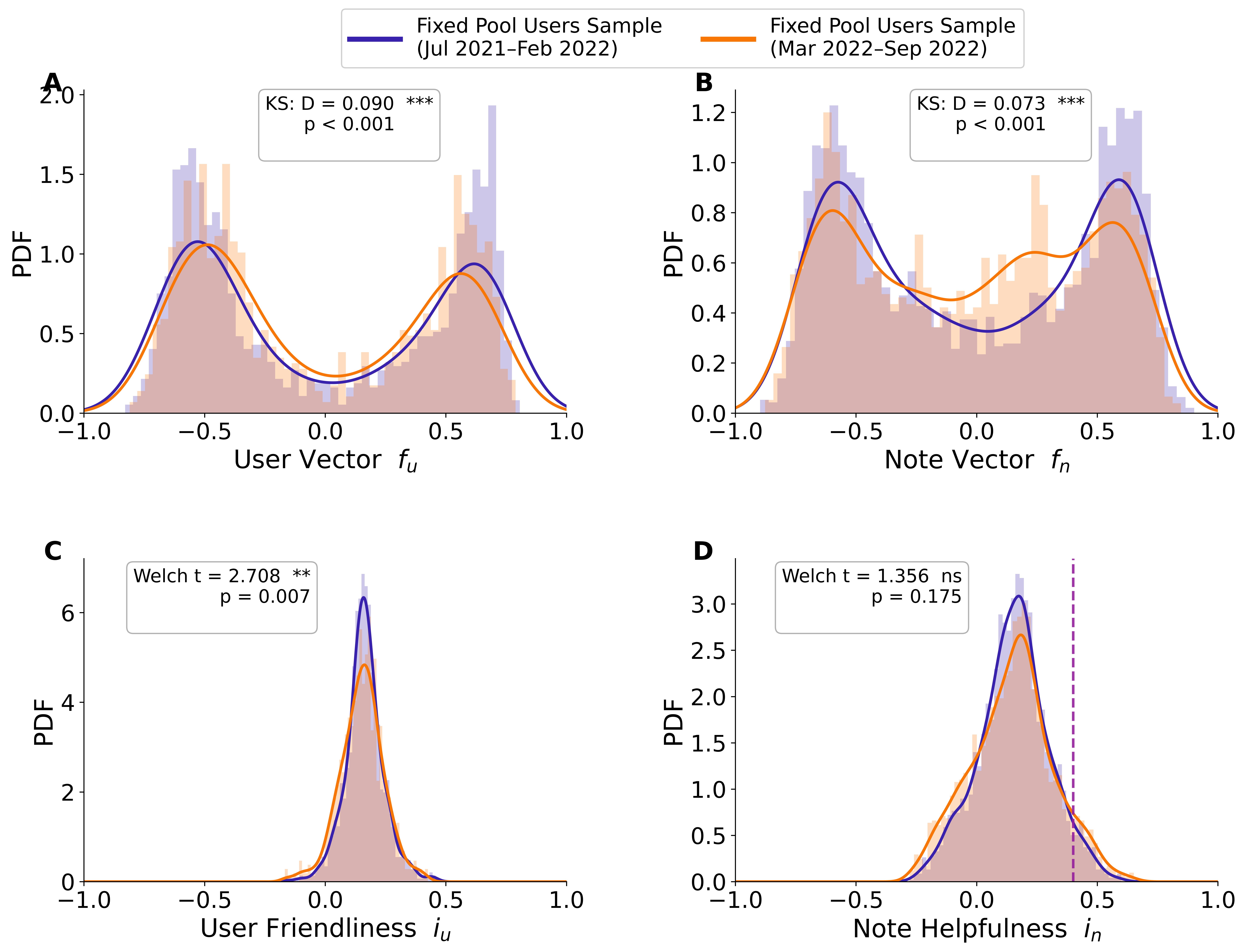}
\caption{\textbf{Estimated distributions of Community Notes Matrix Factorization parameters before and after March 2022 for the Fixed Pool Users Sample} As Figure  \ref{fig:distributions}, restricted to contributors who appear as scored raters in both periods. Panel A: User Vector $f_u$ (KS D = 0.090, p < 0.001). Panel B: Note Vector $f_n$ (KS D = 0.073, p < 0.001). Panel C: User Friendliness $i_u$ (Welch t = 2.708, p = 0.007). Panel D: Note Helpfulness $i_n$ (Welch t = 1.356, p = 0.175); dashed line marks the publication threshold ($i_n \ge 0.40$).}\label{fig:sensitivity_distribution}
\end{figure}

Figure \ref{fig:sensitivity_distribution} compares the distributions of the four Matrix Factorization parameters within this fixed panel, together with the corresponding significance tests. The polarity parameters remain significantly less bimodal after the transition ($f_u$: KS D = 0.090, p $<$ 0.001; $f_n$: KS D = 0.073, p $<$ 0.001), consistent with our main finding of reduced ideological polarization in the algorithm's latent representations. User Friendliness $i_u$ also differs significantly between periods in this subsample (Welch t = 2.708, p = 0.007) — a result that reaches significance here but not in the full-sample comparison (Welch t = 1.280, p = 0.200). We read this as evidence that the same fixed contributors became modestly more selective raters after the transition, a shift partly masked in the full sample by the simultaneous entry of new contributors. Note Helpfulness $i_n$ does not differ significantly between periods (Welch t = 1.356, p = 0.175), matching the full-sample result.

\begin{figure}[H]
\centering
\includegraphics[width=\textwidth]{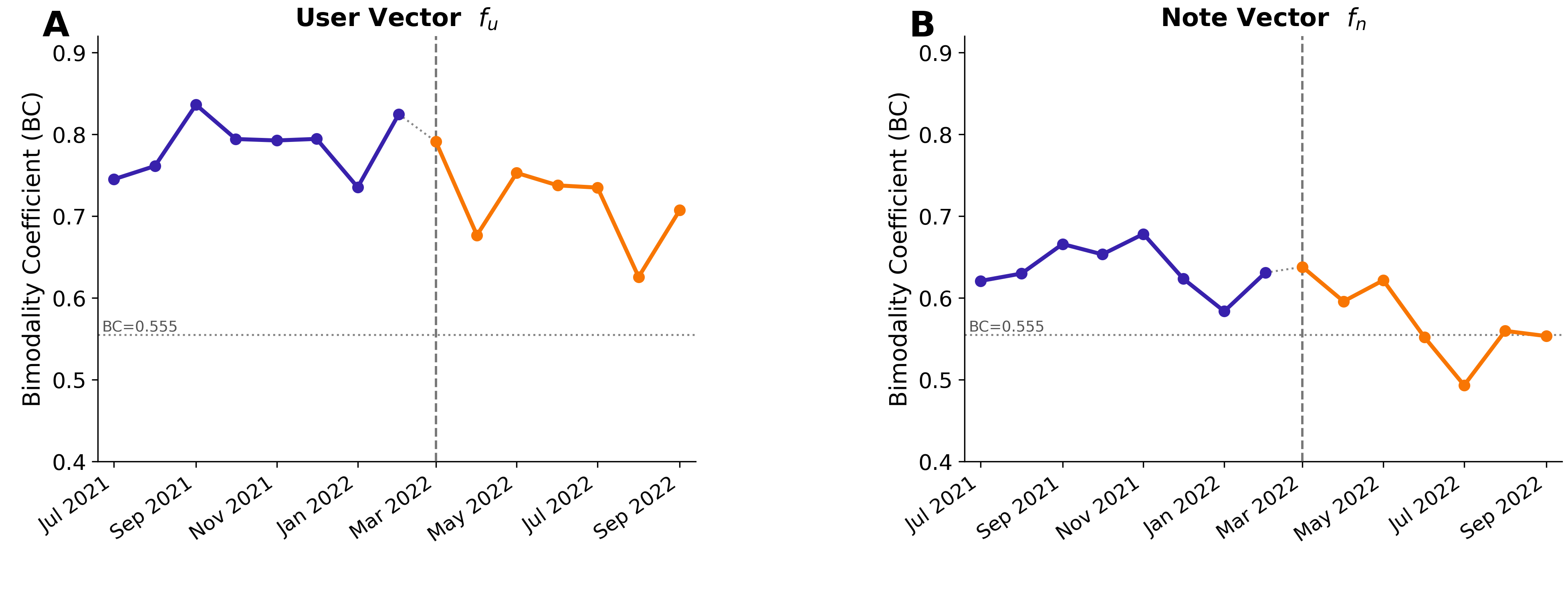}
\caption{\textbf{Monthly evolution of Matrix Factorization parameters for the Fixed Pool Users Sample} As in Figure 2, restricted to contributors who appear as scored raters in both periods. Panel A: Bimodality Coefficient of User Vector $f_u$. Panel B: Bimodality Coefficient of Note Vector $f_n$.}\label{fig:sensitivity_monthly}
\end{figure}

Figure \ref{fig:sensitivity_monthly} shows the monthly evolution of the Bimodality Coefficient of $f_u$ and $f_n$ for the Fixed Pool Users Sample. The trajectories closely reproduce the pattern observed in the full sample: the BC of both $f_u$ and $f_n$ declines after March 2022, approaching or crossing the bimodality threshold (BC = 0.555) after the transition to the new algorithm.

Restricting the comparison to contributors present in both periods therefore reproduces our main finding. Additionally, it suggests that part of the behavioral change we document is attributable to existing contributors adapting their ratings, not solely to compositional turnover in the contributor base.

We applied the same robustness procedure to the network analysis. Specifically, we reconstructed the after network using only contributors present in the first period. This restriction reduced the after network to 1,167 nodes and 3,916 edges.

We then recalculated alignment and antagonism for the restricted networks. The results are consistent with those of the main analysis: alignment decreased (two-sided $z$-test, $z = 230.0$, $P < 0.001$) from $0.740$ (95\% CI [$0.739$, $0.741$]) to $0.620$ (95\% CI [$0.619$, $0.620$]), while antagonism (two-sided $z$-test, $z = -12.06$, $P < 0.001$) increased from $0.320$ (SE = $0.007$) to $0.446$ (SE = $0.008$).

\section{Evaluation Notes}

\begin{figure}[H]
\centering
\includegraphics[width=\textwidth]{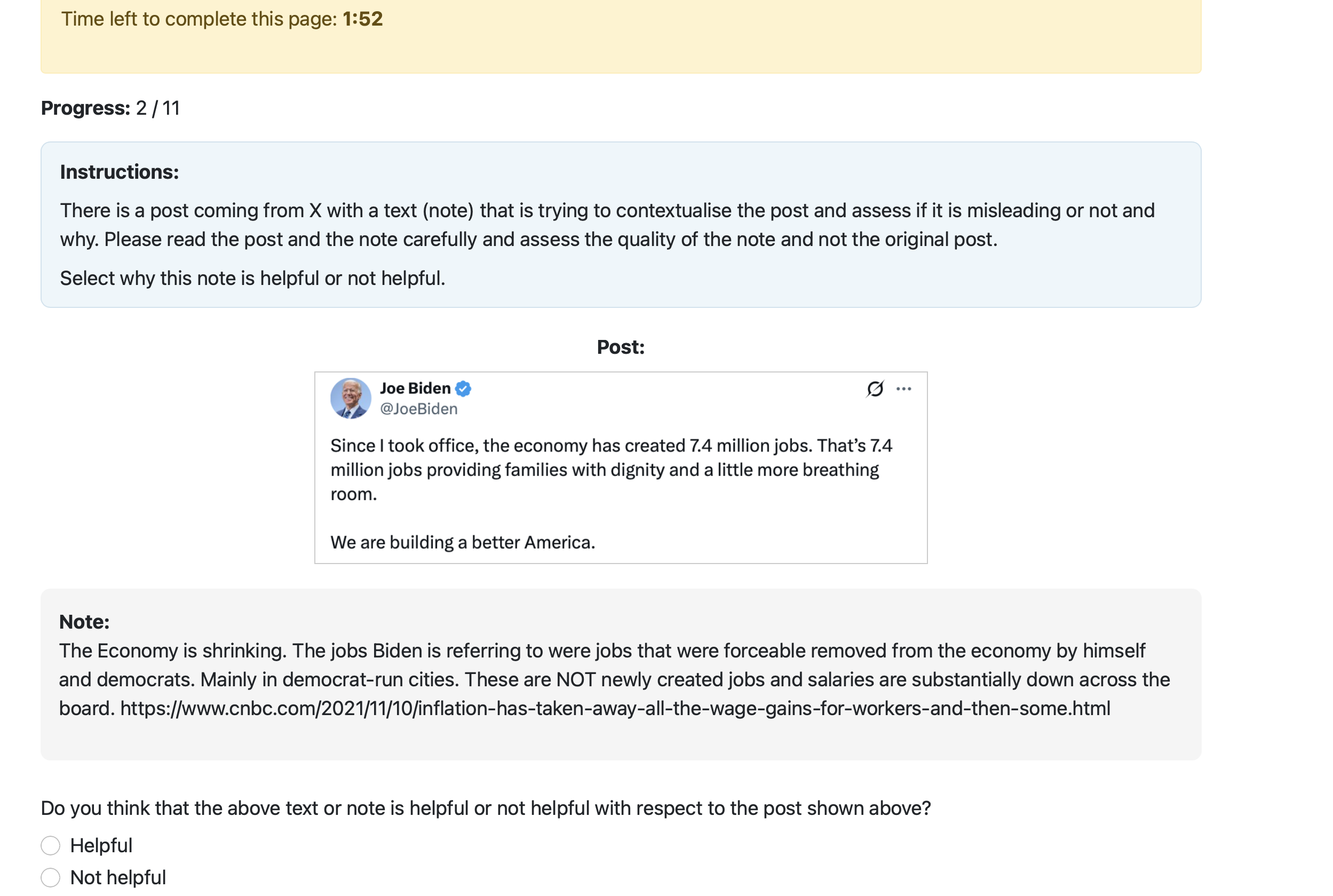}
\caption{\textbf{Example of the independent evaluation task} Screenshot of the oTree interface shown to Prolific participants. Participants viewed the original X post and the corresponding note, judged whether the note was Helpful with respect to the post (Helpful / Not Helpful), and then selected one or more predefined reasons for that judgment from a response list based on their initial answer.}\label{fig:screenshot_experiment}
\end{figure}

Figure \ref{fig:screenshot_experiment} reproduces the participant-facing interface for the online evaluation task. Participants recruited via Prolific were redirected to an experimental interface implemented in oTree, where they viewed a screenshot of the original X post together with the corresponding Community Note and were first asked a binary question — "Do you think that the above text or note is Helpful or not Helpful with respect to the post shown above?" (Helpful / Not Helpful) — with respect to that post. Immediately after this binary judgment, participants answered a follow-up question asking them to select the reason for their assessment. Participants who selected Not Helpful chose among ten predefined categories: sources not included or unreliable, sources do not support the note, incorrect information, opinion or speculation, typos or unclear language, misses key points or is irrelevant, argumentative or biased language, note not needed on this post, harassment or abuse, and other. Participants who selected Helpful chose among six predefined categories: cites high-quality sources, easy to understand, directly addresses the post's claim, provides important context, neutral or unbiased language, and other. This classification represents the reason categories Community Notes itself gives to contributors when they rate notes on the platform, so that our external evaluators assessed helpfulness along dimensions consistent with the platform's own construct of note quality.

\begin{figure}[H]
\centering
\includegraphics[width=0.85\textwidth]{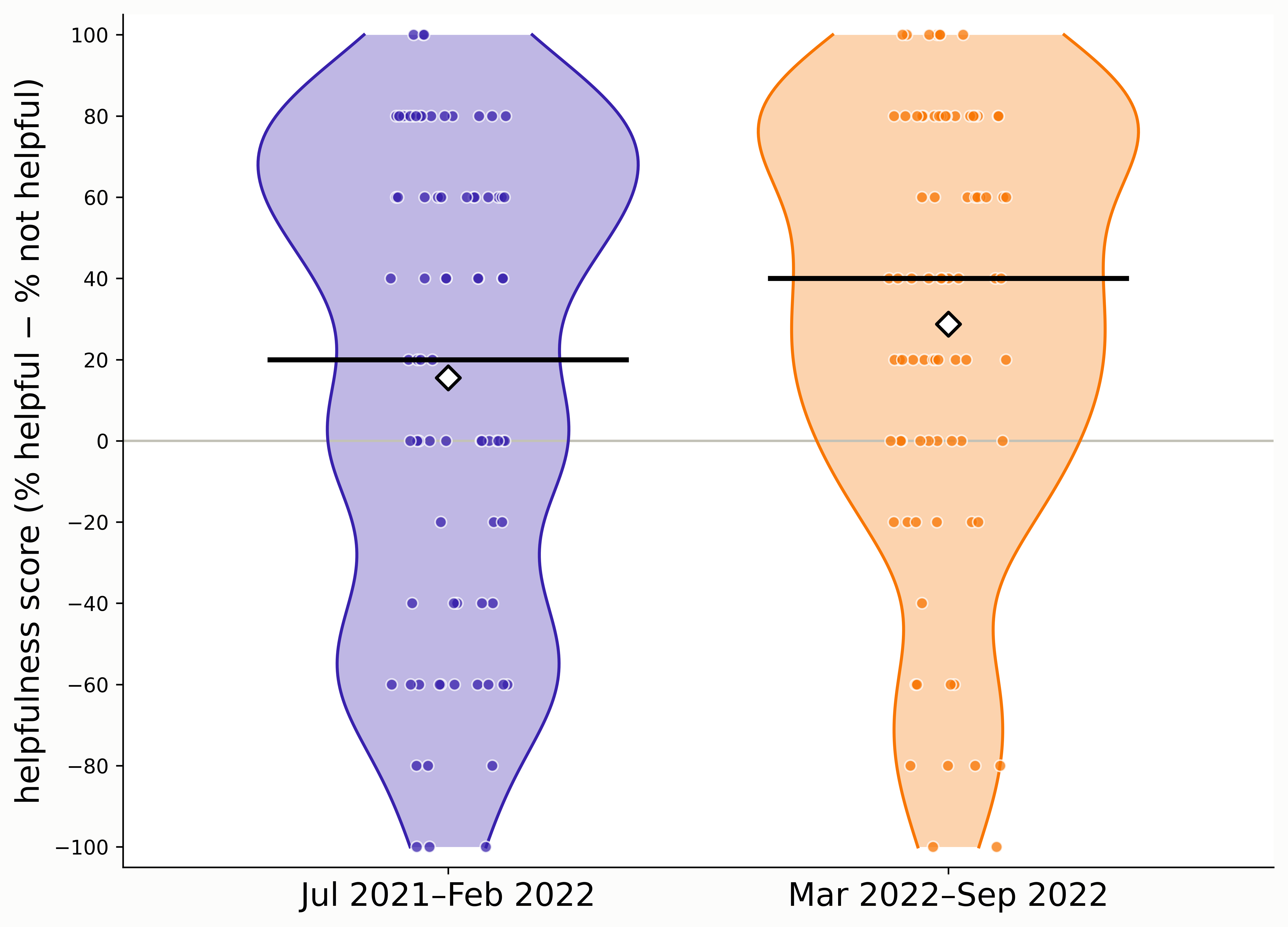}
\caption{\textbf{Distribution of note helpfulness scores before and after the deployment of the bridging algoritm.}
Each note's helpfulness score is the percentage of evaluators who judged the note Helpful minus the percentage who judged it Not Helpful, ranging from $-100$ (unanimously Not Helpful) to $+100$ (unanimously Helpful). Violin plots show the distribution of scores for 80 notes in each period, pooled across two independent rounds of data collection; individual points represent notes, diamonds indicate the mean, and horizontal black lines indicate the median. Notes in the second period had a higher mean helpfulness score than those in the previous one (28.8 vs.\ 15.5) and a higher median (40 vs.\ 20), though the difference was not statistically significant (Mann-Whitney $U = 2786.5$, $p = 0.155$; Cohen's $d = 0.24$).}
\label{fig:violinplot_evaluation}
\end{figure}

\begin{figure}[H]
    \centering
    \includegraphics[width=0.5\linewidth]{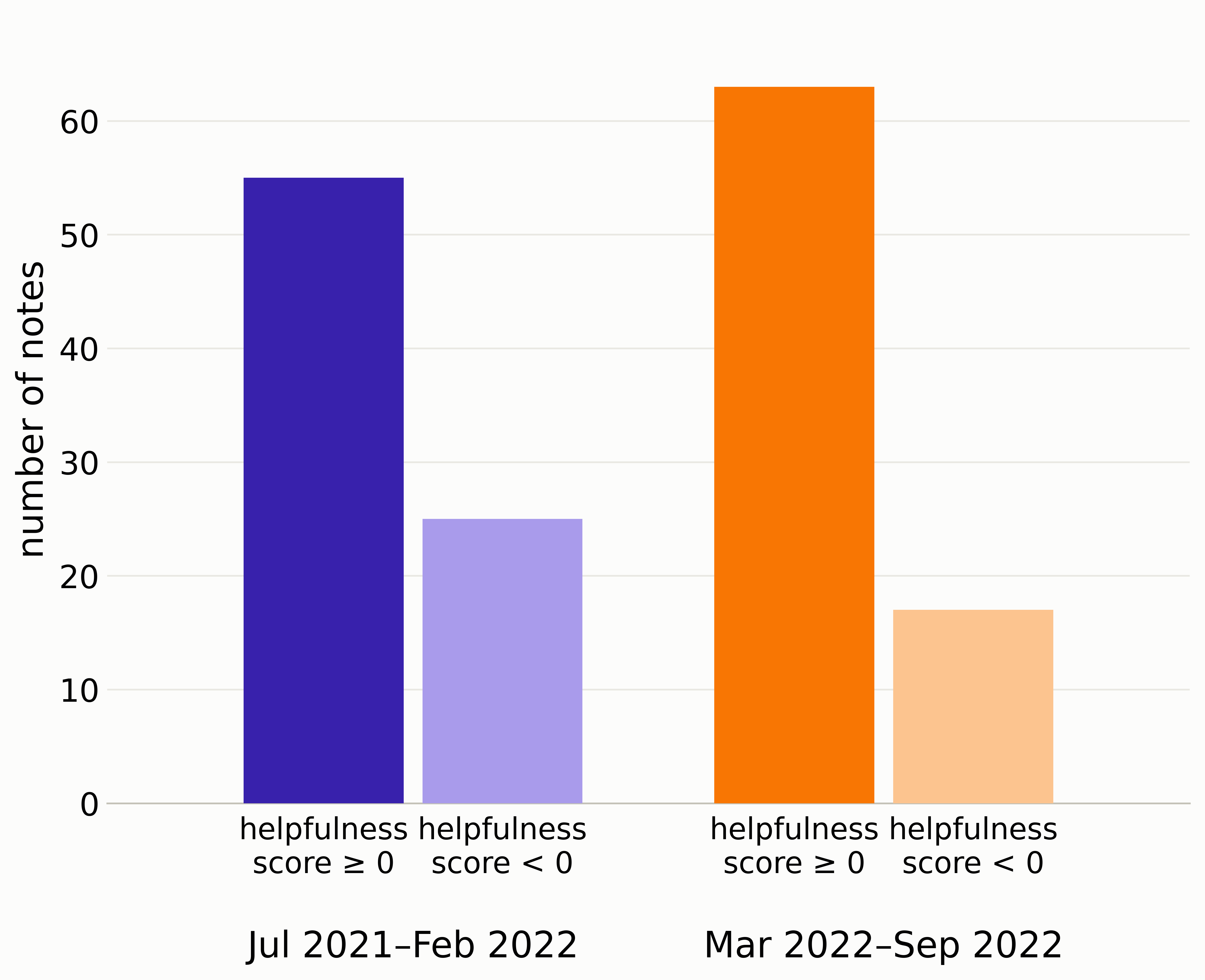}
    \caption{\textbf{Notes judged helpful and not helpful by independent evaluators.}
    Bars show the number of notes with positive (score $\geq 0$) and negative (score $< 0$) helpfulness scores, as assessed by independent evaluators. A note's score is the percentage of evaluators rating it Helpful minus the percentage rating it Not Helpful. After the transition, the number of notes with positive scores increased from 55 to 63, while the number with negative scores decreased from 25 to 17.}
    \label{fig:helpful_counts}
\end{figure}

\section{Sensitivity analysis for the variance threshold}
To construct the signed networks, we applied a variance threshold to determine which edges to retain. For each edge, we estimated the expected probability of Helpfulness ($E_p$) and its posterior variance ($\mathrm{Var}_p$), using the number of Helpful, Somewhat Helpful, and Not Helpful ratings. 

In the main analysis, we used a variance threshold of 0.05. To assess whether this threshold affected the substantive conclusions, we conducted a sensitivity analysis across a range of variance thresholds from 0.005 to 0.10. For each threshold, we reconstructed both networks before and after introducing the MF algorithm and recalculated antagonism and alignment.

\begin{figure}[H]
\centering
\includegraphics[width=\textwidth]{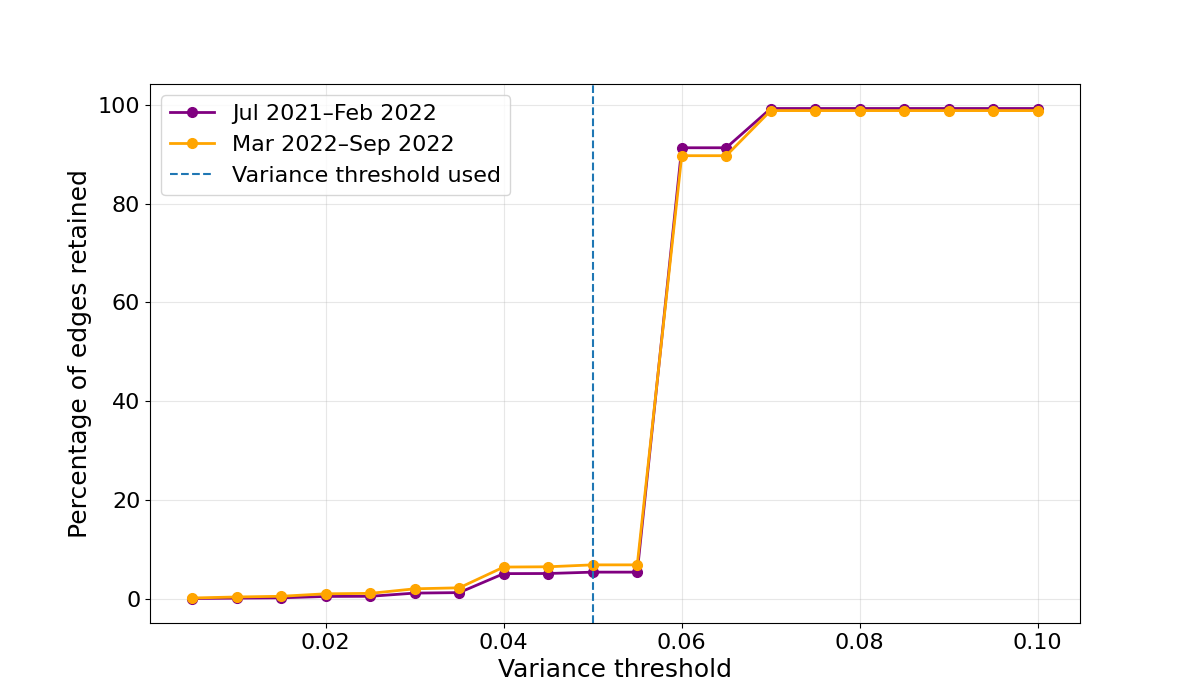}
\caption{\textbf{Sensitivity of edge retention to the variance threshold.} The percentage of edges retained in the before and after networks across different variance thresholds. Edges were retained only when their posterior variance was below the specified threshold. The selected threshold of 0.05 represents the variance threshold used in the main text.}\label{fig:sensitivity_networks}
\end{figure}

The percentage of retained edges varied sharply around the selected threshold. At thresholds below approximately 0.04, fewer than 5\% of edges were retained, indicating that these thresholds were overly restrictive. At thresholds above approximately 0.06, more than 85\% of edges were retained, indicating that these thresholds were comparatively permissive. This pattern was observed for both the before and after networks (Figure~\ref{fig:sensitivity_networks}). Therefore, the threshold of 0.05 provided a reasonable intermediate value: it retained a meaningful subset of edges while still excluding edges with relatively high uncertainty.

\begin{figure}[H]
\centering
\includegraphics[width=0.85\textwidth]{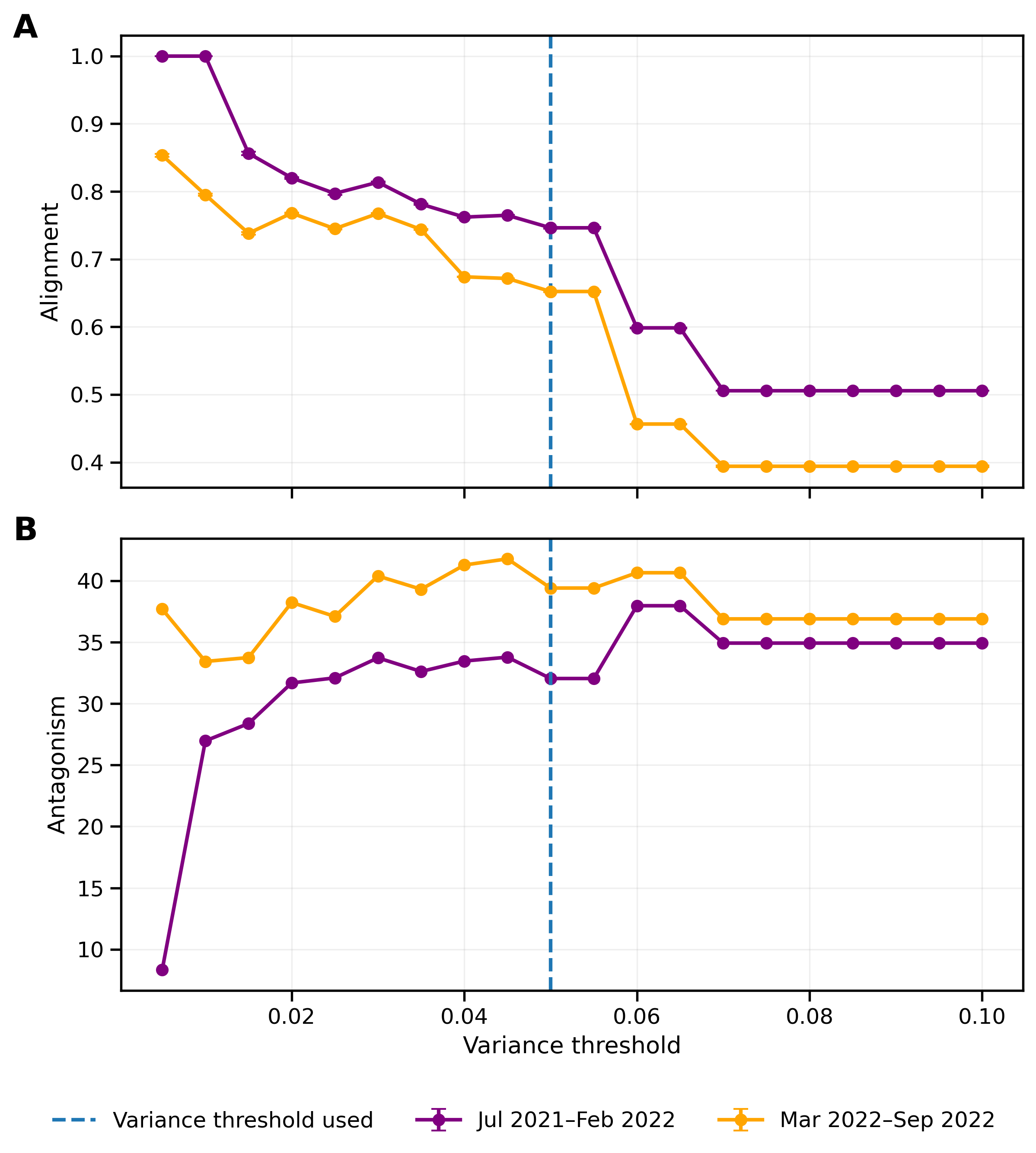}
\caption{\textbf{Sensitivity of alignment and antagonism to the variance threshold.} (A) The antagonism and (B) the alignment of the two networks before and after the introduction of the bridging algorithm across the range of tested variance thresholds. The selected threshold of 0.05 represents the variance threshold used in the main text.}\label{fig:sensitivity_antagonism_alignment}
\end{figure}

The substantive results were robust across the full range of tested thresholds. For all variance thresholds, antagonism is consistently lower in the before network than in the after network (Figure~\ref{fig:sensitivity_antagonism_alignment} (A)). At the same time, alignment is consistently higher in the before network than in the after network (Figure~\ref{fig:sensitivity_antagonism_alignment} (B)). Thus, the direction of the main findings did not depend on the specific variance threshold used to construct the signed networks.

\section{Additional Figures}

\begin{figure}[H]
\centering
\includegraphics[width=0.85\textwidth]{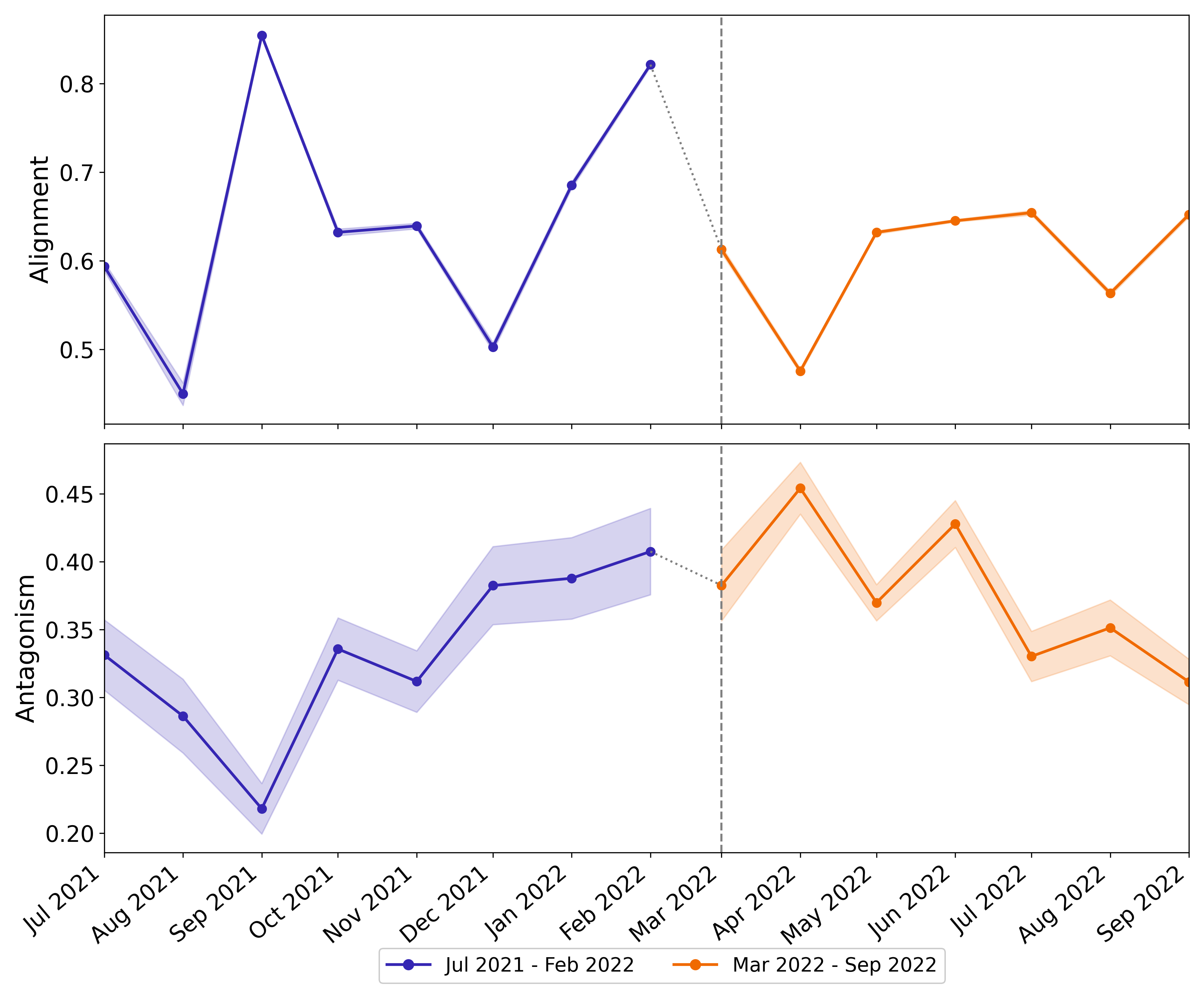}
\caption{\textbf{Temporal Trends in Alignment and Antagonism.}
Panel (A): Monthly estimates of alignment and Panel (B): Monthly estimates of antagonism across the two periods. Shaded regions for alignment represent 95\% confidence intervals, while shaded regions for antagonism represent standard errors. The vertical dashed line marks the transition from before the deployment of the bridging algorithm to after that in March 2022.}
\label{fig:alignment_antagonism_monthly}
\end{figure}
\end{document}